\RequirePackage{fix-cm}
\documentclass[smallextended]{svjour3}       
\smartqed  
\usepackage{graphicx}
\usepackage{amsmath}
\usepackage{amssymb}
\usepackage{subfig}
\journalname{J Low Temp Phys}
\begin{document}

\title{Remarks on Thermodynamic Properties of a Double Ring-Shaped Quantum Dot at Low and High Temperatures
}

\titlerunning{Remarks on Themodynamics Properties}        

\author{ Andr\'{e}s G. Jir\'{o}n Vicente \and
Luis B. Castro         \and
        Angel E. Obispo \and
        Luis E. Arroyo Meza 
}


\institute{Luis B. Castro \at
              Departamento de F\'{\i}sica e Qu\'{\i}mica, Universidade Estadual Paulista (UNESP), Campus de Guaratin\-gue\-t\'{a}, 12516-410, Guaratinguet\'{a}, SP, Brazil. \\
              \email{luis.castro@ufma.br}           
           \and
           Andr\'{e}s G. Jir\'{o}n Vicente \and Luis B. Castro \and Angel E. Obispo \and Luis E. Arroyo Meza\at 
           Departamento de F\'{\i}sica, Universidade Federal do Maranh\~{a}o, Campus Universit\'{a}rio do Bacanga, 65080-805, S\~{a}o Lu\'{\i}s, MA, Brazil.                  
}

\date{Received: date / Accepted: date}

\maketitle

\begin{abstract}

In a recent paper published in this Journal, Khordad and collaborators [J Low Temp Phys (2018) 190:200] have studied the thermodynamics properties of a GaAs double ring-shaped quantum dot under external magnetic and electric fields. In that meritorious research the energy of system was obtained by solving the Schr\"{o}dinger equation. The radial equation was mapped into a confluent hypergeometric differential equation and the differential equation associated to $z$ coordinate was mapped into a biconfluent Heun differential equation. In this paper, it is pointed out a misleading treatment on the solution of the biconfluent Heun equation. It is shown that the energy $E_{z}$ can not be labeled with $n_{z}$ and this fact jeopardizes the results of this system. We calculate the partition function with the correct energy spectrum and recalculate the specific heat and entropy as a function of low and high temperatures.

\end{abstract}

\section{Introduction}
\label{intro}

In the past thirty years, semiconductor quantum dots were the subject of great experimental \cite{RMP64:849:1992,JCP112:7790:2000,EL32:1732:1996} and theoretical \cite{CEJP6:97:2008,JAP112:083514:2012,PRB46:3898:1992,JAP88:730:2000,PLA134:395:1989,JLTP190:200:2018,%
SLM110:146:2017,JPCM7:965:1995} interest, mainly because they offer significantly improved electronic and optical properties associated with the quantum confinement in all three spatial dimensions of a few electrons at the semiconductor interface to form quase-zero-dimensional systems ($10-1000$\,A length scale). As the electrons wavelength is of the same length scale as the confinement, the quantum effects become relevant being the most notable the emergence of a quantized energy spectrum with spacing of a few meV. This, along with other remarkable properties, increased the potential applications of quantum dots, for instance, in micro-electronic devices such as quantum dot lasers \cite{JPCM15:R1063:2003,APL70:981:1997}, solar cells \cite{SE85:1264:2011,PPRA10:433:2002}, single electron transistors \cite{S320:356:2008,PRB57:15400:1998} and quantum computers \cite{PRB61:13813:2000,CPC141:66:2001}. For more details, the interested reader will find more information in \cite{RPP64:701:2001,RMP74:1283:2002,RMP75:1:2002}.

From a theoretical point of view, to study quantum dots it is essential to know the profile of the confining potential which often is represented like a symmetrical paraboidal potential \cite{CEJP6:97:2008}, a Gaussian confinement \cite{JAP112:083514:2012}, a spherical harmonic oscillator \cite{PRB46:3898:1992}, a pyramidal potential \cite{JAP88:730:2000}, a ring-shaped oscillator \cite{PLA134:395:1989}, or a double ring-shaped oscillator \cite{JLTP190:200:2018,SLM110:146:2017}. This last case was addressed in a recent paper published in this Journal by Khordad et al. \cite{JLTP190:200:2018}. Particularly, the authors studied the thermodynamic properties of GaAs double ring-shaped quantum dot under external magnetic and electric fields by means of a theoretical model and analytical solutions. To achieve their goal, the authors need to calculate the energy spectrum, which has been obtained by solving the Schr\"{o}dinger equation with a confining potential, constant magnetic and electric fields \cite{JPCM7:965:1995}. Considering cylindrical coordinates, the radial differential equation was mapped into a confluent hypergeometric differential equation. On the other hand, the differential equation related to z axis was mapped into a biconfluent Heun differential equation. 

The purpose of this paper is point to out a misleading treatment on the solution of the biconfluent Heun equation, this fact jeopardizes the results obtained in \cite{JLTP190:200:2018}, because the thermodynamic properties of the system depend mainly on the energy spectrum. With the energy spectrum, the partition function is properly built and two thermodynamic properties: the specific heat and entropy are recalculated as a function of temperature. Finally, our results are compared with those found in the literature and, in special, with Ref. \cite{JLTP190:200:2018}.

\section{Theory and model}

The Hamiltonian of an electron confined in a double ring-shaped quantum dot under magnetic and electric fields is given by \cite{JLTP190:200:2018}
\begin{equation}\label{hamil}
H=\frac{\left( \vec{P}+e\vec{A} \right)^{2}}{2m^{\ast}}+V(r,\theta)-e\vec{F}\cdot\vec{r}\,,
\end{equation}
\noindent where
\begin{equation}\label{vcon}
V(r,\theta)=\frac{1}{2}m^{\ast}\omega_{0}^{2}r^{2}+\frac{\hbar^{2}}{2m^{\ast}}\left( \frac{\bar{B}}{r^{2}\sin^{2}\theta}+
\frac{\bar{C}}{r^{2}\cos^{2}\theta} \right)\,,
\end{equation}
\noindent $m^{\ast}$ is the effective mass of the electron, $\vec{A}=\frac{B}{2}\left( -y,x,0 \right)$ and $\vec{F}$ is the electric field. Here $\bar{B}$ and $\bar{C}$ are two potential parameters.

In cylindrical coordinates, the resulting stationary Schr\"{o}dinger equation (\ref{hamil}) is
\begin{equation}\label{eq2o}
\begin{split}
& \left[ -\frac{\hbar^{2}}{2m^{\ast}}\nabla^{2}+ \frac{1}{8}m^{\ast}\omega_{c}^2\rho^{2}+\frac{1}{2}\omega_{c}L_{z}
+\frac{1}{2}m^{\ast}\omega_{0}^{2}(\rho^{2}+z^{2}) \right.\\
& \left. +\frac{\hbar^{2}}{2m^{\ast}}\left( \frac{\bar{B}}{\rho^{2}}+\frac{\bar{C}}{z^{2}}-eF_{z}z \right) \right]\psi(\rho,\phi,z)=E\psi(\rho,\phi,z)\,.
\end{split}
\end{equation}
\noindent At this stage, let us consider the variable separation method and using the solution of (\ref{eq2o}) in the form of $\psi(\rho,\phi,z)=R(\rho)F(z)\mathrm{e}^{im\phi}$, one finds the following equations
\begin{equation}\label{eq2or}
\begin{split}
\frac{d^{2}R(\rho)}{d\rho^{2}}+\frac{1}{\rho}\frac{dR(\rho)}{d\rho}& -\frac{m^{2}+\bar{B}}{\rho^{2}}R(\rho)
-\frac{(m^{\ast})^{2}\Omega^{2}}{\hbar^{2}}\rho^{2}R(\rho)\\
& +\left( \frac{2m^{\ast}E_{\rho}}{\hbar^{2}}-\frac{m^{\ast}m\omega_{c}}{\hbar} \right)R(\rho)=0\,,
\end{split}
\end{equation}
\noindent and
\begin{equation}\label{eq2oz}
\frac{d^{2}F(z)}{dz^{2}}-\left( \frac{(m^{\ast})^{2}\omega_{0}^{2}}{\hbar^{2}}z^{2}+\frac{\bar{C}}{z^{2}}-
\frac{2m^{\ast}E_{z}}{\hbar^{2}}-\frac{2m^{\ast}}{\hbar^{2}}eF_{z}z \right)F(z)=0\,,
\end{equation}
\noindent where $\omega_{c}=eB/m^{\ast}$ is the cyclotron frequency and $\Omega=\sqrt{\omega_{0}^{2}+\frac{\omega_{c}^{2}}{4}}$\,.

\subsection{Radial equation}

Making use of the new variable $\eta=\kappa\rho^2$ with $\kappa=\frac{m^{\ast}\Omega}{\hbar}$, the Eq. (\ref{eq2or}) becomes
\begin{equation}\label{eq2orr2}
\left(\frac{d^2 }{d\eta^2}+\frac{1}{\eta}\frac{d}{d\eta}-\frac{m^2+\bar{B}}{4\eta^2}+\frac{\lambda}{4\kappa\eta}-\frac{1}{4}\right)R(\eta)=0\,,
\end{equation}
\noindent where $\lambda=\frac{2m^{\ast}E_{\rho}}{\hbar^2}-\frac{m^{\ast}m\omega_{c}}{\hbar}$. The solution for all $\eta$ can be expressed as
\begin{equation}\label{ansatz1}
R(\eta)=\eta^{\frac{|\sqrt{m^2+\bar{B}}|}{2}}\mathrm{e}^{-\frac{\eta}{2}}f(\eta)\,,
\end{equation}
\noindent subsequently, by introducing the following parameters
\begin{eqnarray}
a &=& \frac{1}{2}\left( |\sqrt{m^2+\bar{B}}|+1-\frac{\lambda}{2\kappa} \right)\,,\label{a}\\
b &=& |\sqrt{m^2+\bar{B}}|+1\,,\label{b}
\end{eqnarray}
\noindent one finds that $f(\eta)$ can be expressed as a regular solution of the confluent hypergeometric equation (Kummer's function) \cite{ABRAMOWITZ1965}
\begin{equation}\label{hyperg1}
\eta\frac{d^{2}f}{d\eta^2}+(b-\eta)\frac{df}{d\eta}-af=0\,.
\end{equation}
\noindent The solution of Eq. (\ref{hyperg1}), regular at $\eta=0$, can be obtained considering the Kummer's function denoted by $f(\eta)=M(a,b;\eta)$. An important feature of the Kummer's function related to its asymptotic behavior demands that the parameter $a$ to be a non-negative integer. This is a well known condition that permit us obtain the energy spectrum of this system. The quantization condition ($a=-n_{\rho}$) implies into
\begin{equation}\label{erho}
E_{\rho}=\hbar\Omega\left( 2n_{\rho}+1+\sqrt{m^2+\bar{B}} \right)+\frac{m\hbar\omega_{c}}{2}\,.
\end{equation}
\noindent The energy $E_{\rho}$  is in agreement with Ref. \cite{JLTP190:200:2018}. 

\subsection{$z$-axis equation}

Considering the solution of (\ref{eq2oz}) as 
\begin{equation}\label{ansatz}
F(z)=z^{\frac{1+\sqrt{1+4\bar{C}}}{2}}\exp\left(-\frac{m^{\ast}\omega_{0}}{2\hbar}z^{2}\right)\exp\left( \frac{eF_{z}}{\hbar\omega_{0}}z \right)g(z)\,,
\end{equation}
\noindent and defining the new variable $z\rightarrow \sqrt{\frac{m^{\ast}\omega_{0}}{\hbar}}z$, one finds that $g(z)$ can be expressed as a solution of the biconfluent Heun differential equation \cite{JPA19:3527:1986,RONVEAUX1995,PRC86:052201:2012,EPJC72:2051:2012,PLA376:2838:2012,AP341:86:2014,AP355:48:2015,EPJC78:494:2018}
\begin{equation}\label{eqbh}
\frac{d^{2}g(z)}{dz^{2}}+\left(\frac{\alpha+1}{z}-\beta-2z \right)\frac{dg(z)}{dz}+
\left( \gamma-\alpha-2-\frac{\Theta}{z} \right)g(z)=0\,,
\end{equation}
\noindent where 
\begin{eqnarray}
\alpha &=& \sqrt{1+4\bar{C}}\,,\label{alpha}  \\
\beta &=& -\frac{2eF_{z}}{\hbar\omega_{0}}\sqrt{\frac{\hbar}{m^{\ast}\omega_{0}}}\,,\label{beta}\\
\gamma &=& \frac{2E_{z}}{\hbar\omega_{0}}+\frac{e^{2}F_{z}^{2}}{m^{\ast}\hbar \omega_{0}^{3}}\,,\label{gamma}\\
\Theta &=& \frac{1}{2}\left[ \delta+\beta(1+\alpha) \right]\,,\label{Theta}
\end{eqnarray}
\noindent with $\delta=0$. Note that the expressions for $\beta$ and $\gamma$ obtained in Ref. \cite{JLTP190:200:2018} are different to our results, probably due to erroneous calculations in the manipulation of the equation (\ref{eq2oz}).  In this ways, the regular solution of (\ref{eqbh}) is given by
\begin{equation}\label{solution}
H_{b}(\alpha,\beta,\gamma,0;z)=\sum_{j=0}\frac{\Gamma(1+\alpha)}{\Gamma(1+\alpha+j)}a_{j}\frac{z^{j}}{j!}\,,
\end{equation}
\noindent where $\Gamma(x)$ is the gamma function, $a_{0}=1$ and $a_{1}=\Theta$. The remaining coefficients for $\beta\neq0$ satisfy the recurrence relation,
\begin{equation}\label{rr}
a_{j+2}=\left[(j+1)\beta+\Theta\right]a_{j+1}-(j+1)(j+1+\alpha)(\Delta-2j)a_{j}\,,\quad j\geq0\,,
\end{equation}
\noindent where $\Delta=\gamma-\alpha-2$. Using (\ref{rr}) for $j=0$, we can obtain $a_{2}=(\beta+\Theta)\Theta-(\alpha+1)\Delta$ and so on, gets the others coefficients for $j>0$. From recurrence relation (\ref{rr}), $H_{b}$ becomes a polynomial of degree $n$ if only if two conditions are satisfied \cite{RONVEAUX1995}
\begin{eqnarray}
\Delta &=& 2n_{z}\,,\quad (n_{z}=0,1,\ldots)\label{cc1} \\
a_{n_{z}+1} &=& 0 \,.\label{cc2}
\end{eqnarray}  
\noindent At this stage, it is worthwhile to mention that the energy of the system is obtained using both conditions (\ref{cc1}) and (\ref{cc2}).

From the condition (\ref{cc1}), one obtains
\begin{equation}\label{enern}
E_{z,n_{z}}=\left( n_{z}+1+\frac{\alpha}{2}-\frac{e^{2}F_{z}^{2}}{2m^{\ast}\hbar\omega^{3}} \right)\hbar\omega_{0}\,.
\end{equation}
The problem does not end here, it is necessary to analyze the second condition of quantization. Now we focus attention on the condition (\ref{cc2}), this condition provides a constraint on the values of potential parameters. For instance, $n_{z}=0$ ($\Delta=0$) implies that $a_{1}=\Theta=0$. This condition furnishes the following equation
\begin{equation}\label{cc2n0}
\beta(1+\alpha)=0\,.
\end{equation} 
\noindent One finds that is not possible to extract a physically acceptable expression for $F_{z}$ or $\omega_{0}$ from (\ref{cc2n0}). Thus, $n_{z}=0$ is not an allowed value. 

Now, let us consider the case $n_{z}=1$ ($\Delta=2$), which implies that $a_{2}=(\beta+\Theta)\Theta-2(1+\alpha)=0$. This condition for $\beta<0$ and $F_{z}>0$ or $\beta>0$ and $F_{z}<0$ provides the following constrain 
\begin{equation}\label{vn1}
\frac{e^{2}F_{z,1}^{2}}{2m^{\ast}\hbar\omega_{0}^{3}}=\frac{1}{\alpha+3}\,.
\end{equation}
\noindent Substituting (\ref{vn1}) into (\ref{enern}) for $n_{z}=1$, we find
\begin{equation}\label{enern1}
E_{z,1}=\left( \frac{\sqrt{1+4\bar{C}}}{2}+2-\frac{1}{\sqrt{1+4\bar{C}}+3} \right)\hbar\omega_{0}\,.
\end{equation}
\noindent The last expression represents the energy of the system for $n_{z}=1$. 

Following the same procedure, we can obtain the energy of the system for $n=2$ ($\Delta=4$), considering the constrain on the values of potential parameters from the condition $a_{3}=0$. Explicitly, for $\beta<0$ and $F_{z}>0$ or 
$\beta>0$ and $F_{z}<0$ we obtain 
\begin{equation}\label{vn2}
\frac{e^{2}F_{z,2}^{2}}{2m^{\ast}\hbar\omega_{0}^{3}}=\frac{2(2\alpha+7)}{\alpha^{2}+8\alpha+15}\,.
\end{equation}
\noindent One more time, substituting (\ref{vn2}) into (\ref{enern}) for $n_{z}=2$, we find
\begin{equation}\label{enern2}
E_{z,2}=\left( \frac{\sqrt{1+4\bar{C}}}{2}+3-\frac{2\sqrt{1+4\bar{C}}+7}{4\sqrt{1+4\bar{C}}+2\bar{C}+8} \right)\hbar\omega_{0}\,.
\end{equation}
\noindent The last expression represents the energy of the system for $n_{z}=2$. For $n_{z}\geq3$, the form of the constraint becomes increasingly cumbersome. With this result, we conclude that the energy $E_{z}$ can not be labeled with $n_{z}$, because both constrain and energy are different for each value of $n_{z}$. This is a peculiar behavior of the biconfluent Heun equation.

\section{Thermodynamic properties}

In order to calculate thermodynamic properties of the system, we need to build the partition function. As it is known, the partition function can be calculated by direct summation over all possible states available to the system. The total energy of the system is given by
\begin{equation}\label{enert}
\begin{split}
E_{n_{\rho},m,1}=& \hbar\Omega\left( 2n_{\rho}+1+\sqrt{m^2+B} \right)+\frac{m\hbar\omega_{c}}{2}\\
& +\left( \frac{\sqrt{1+4C}}{2}+2-\frac{1}{\sqrt{1+4C}+3} \right)\hbar\omega_{0}\,,
\end{split}
\end{equation}   
\noindent for $n_{z}=1$ and satisfying the constraint (\ref{vn1}). Note that each value of $n_{z}$ is associated to a different physical scenario, for this reason we have to fix the value of $n_{z}$.

The partition function is given by (considering $n_{z}=1$)
\begin{equation}\label{fp}
Q=\sum_{n_{\rho}}\sum_{m}\exp{\left( -\beta E_{n_{\rho},m,1} \right)}\,,
\end{equation}
\noindent where $\beta=\frac{1}{k_{B}T}$ is the Boltzmann constant and $T$ is the temperature. To show our results and compare with Ref. \cite{JLTP190:200:2018}, let's recalculate two thermodynamic properties of this system using the  following relations
\begin{eqnarray}
\mathrm{Specific \,\, heat}&:&\quad C_{v} = \frac{\partial U}{\partial T}\,,\label{spheat}\\
\mathrm{Entropy}&:&\quad S = k_{B}\ln Q-k_{B}\beta\frac{\partial\ln Q}{\partial\beta}\,.\label{entro}
\end{eqnarray}

\section{Results and discussions}

\begin{figure}[!htb]
\centering
\subfloat[]{
\includegraphics[height=6.6cm]{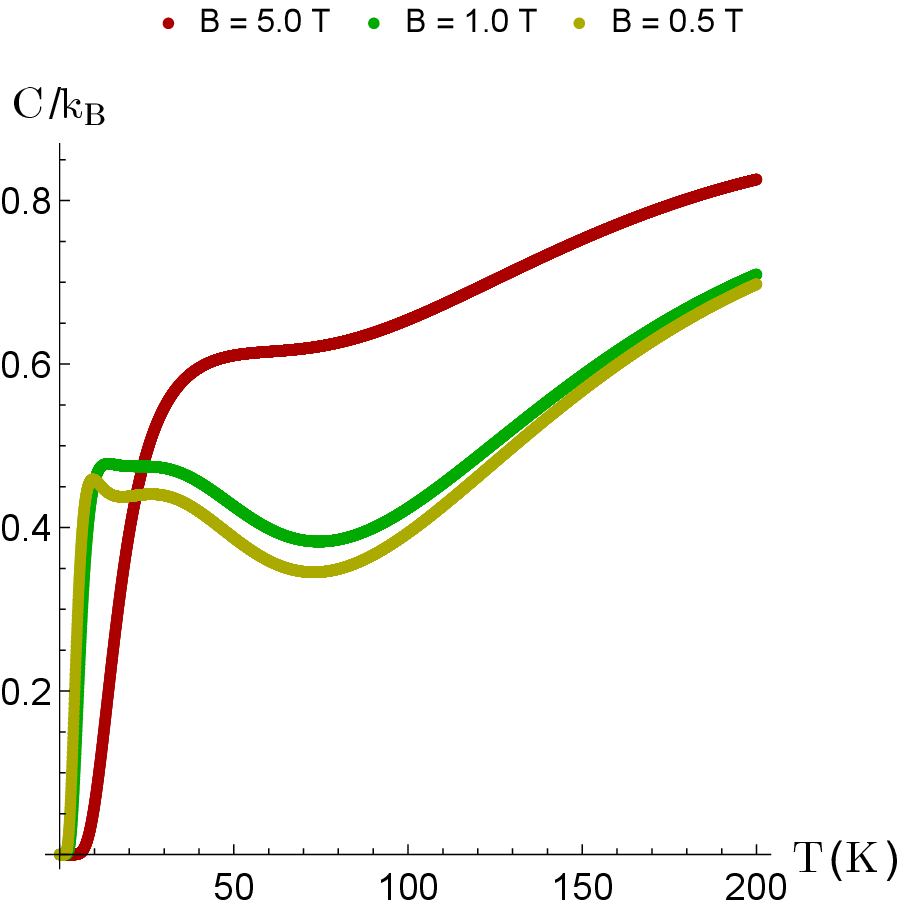}
\label{ccat}
}
\quad 
\subfloat[]{
\includegraphics[height=6cm]{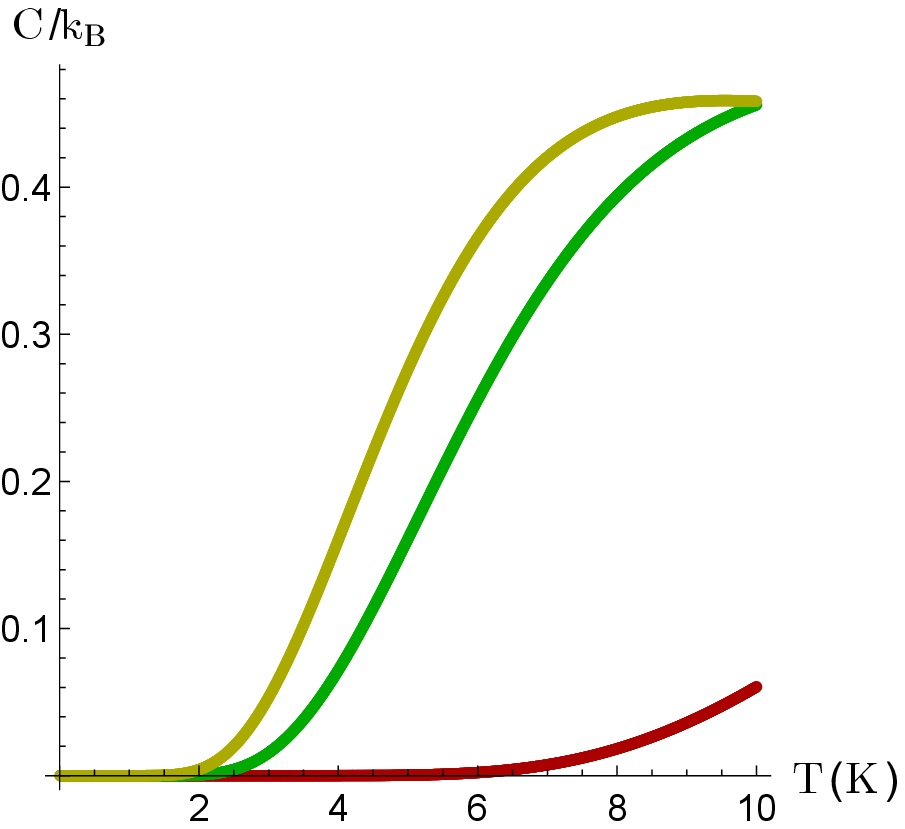}
\label{ccbt}
}
\caption{Specific heat as a function of temperature for different magnetic fields. The curves in (a) and (b) correspond to high and low temperatures, respectively.}
\label{fig:cc}
\end{figure}

\begin{figure}[!htb]
\centering
\subfloat[]{
\includegraphics[height=6.6cm]{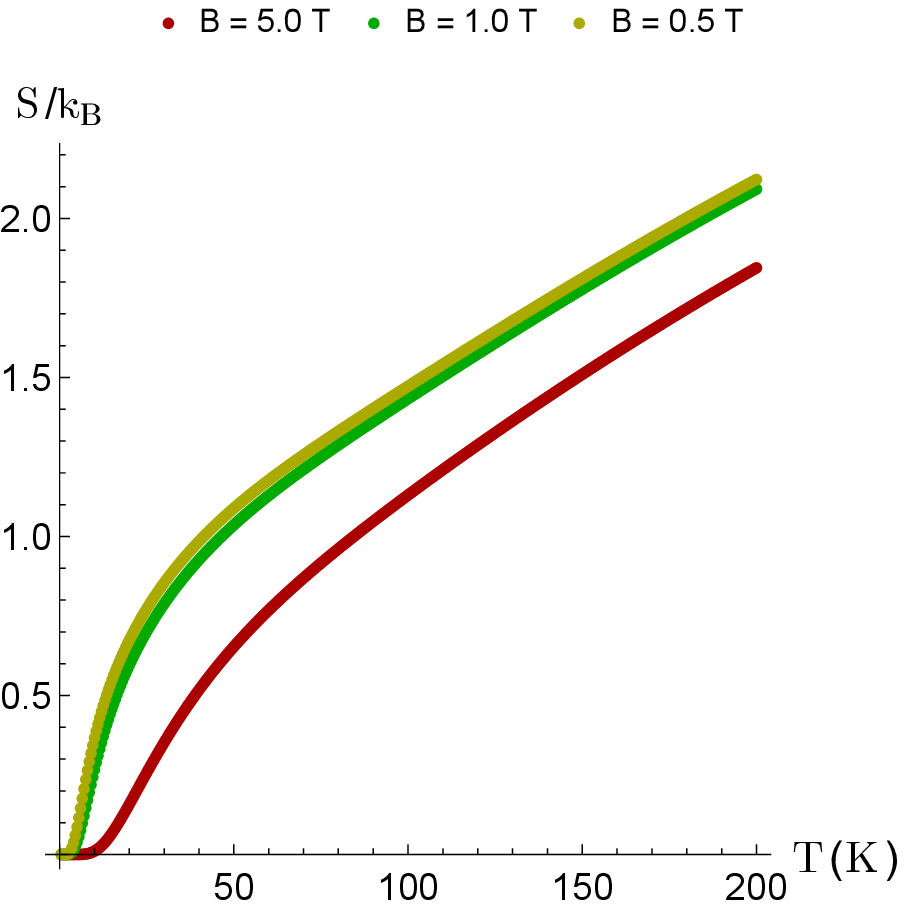}
\label{eat}
}
\quad 
\subfloat[]{
\includegraphics[height=6cm]{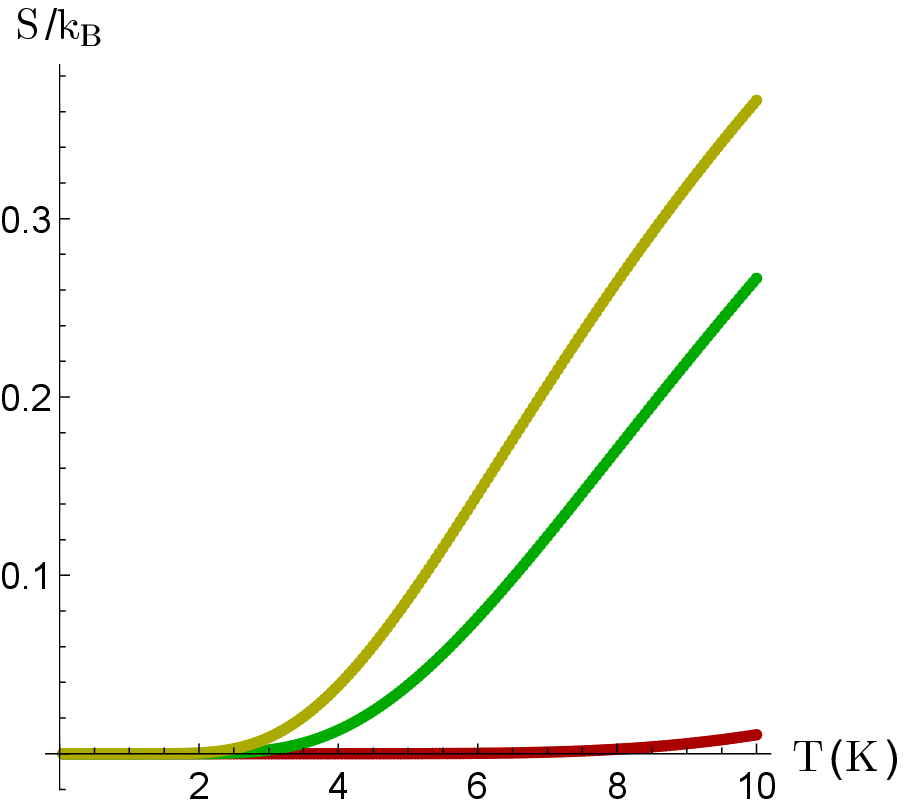}
\label{ebt}
}
\caption{Entropy as a function of temperature for different magnetic fields. The curves in (a) and (b) correspond to high and low temperatures, respectively.}
\label{fig:e}
\end{figure}

Figure \ref{fig:cc} illustrates the behavior of the specific heat of a GaAs QD as a function of the temperature ($T$) for different magnetic fields: $5.0$, $1.0$ and $0.5$ teslas. Note that at low temperatures (Fig. \ref{ccbt}) the magnetic field has no effect on the specific heat ($T<2$\,K). For $T>2$\,K, the specific heat changes abruptly, increasing with enhancing the temperature and at a fixed temperature, it decreases with enhancing the magnetic field. It is worthwhile to mention that this behavior of the specific heat at a comparatively small temperature range ($T < 10$\,K) agrees with Ref. \cite{JLTP190:200:2018}. On the other hand, from Fig. \ref{ccat}, it is observed that the specific heat increases until a peak structure (resonance), which becomes more pronounced as magnetic fields gets weaker, while it shifts towards left to lower temperature values. This is the well-known Schottky anomaly \cite{JPCS9:285:1959}, which is often observed in magnetic systems and is closely related to the energy required for a thermal transition between the ground state and the first excited state of the system. For our GaAs QD model, such energy spacing is approximately $\Delta E\approx 40$\,meV which correspond to radio frequencies of $\nu\sim 10^{11}$\,Hz, and whose wavelengths are in the infrared region for GaAs. Based on this, we believe our results, together with the observation of the Schottky anomaly, would find experimental realization in the so-called quantum dots photodetectors \cite{NT29:124003:2018,APL78:2428:2001} and photodiodes \cite{APL99:031102:2011,N418:612:2002}. Additionally, we also found a progressive disappearance of the Schottky anomaly as the magnetic field and/or temperature increases and transforming the peak into a shoulder, as shown in Fig. \ref{ccat}. Note that at high temperatures ($T \sim 80$\,K) the specific heat continues to increase with enhancing the temperature, until approaching to some finite value $\sim 2k_B$ (though not shown in the figure) as would be expected from a Dulong-Petit-like behavior. These latest results show a behavior contrary to Ref. \cite{JLTP190:200:2018} but they agree with Refs. \cite{JAP112:083514:2012,PRB54:14532:1996,PA548:123871:2020}, in which the expected behavior for GaAs QD model is shown.

In Figure \ref{fig:e}, we plot the entropy as a function of the temperature (T), for $B=0.5$, $1.0$ and $5.0$ teslas. As expected, clearly it is shown that the entropy increases with enhancing the temperature at a fixed value of the magnetic field. However, the behavior is qualitatively different at distinct temperature regimes. For example, at low temperatures (Fig. \ref{ebt}) the magnetic field has no effect on the entropy ($T<2$\,K). For $T>2$\,K, the entropy is found to be sensitive to magnetic field, in such a way that it increases quite rapidly as the magnetic field decreases. On the other hand, at higher temperatures (Fig. \ref{eat}), entropy increases monotonically until attains a saturation value independent of magnetic field $\sim 7k_B$, i.e., at room temperature, the thermal energy dominates over magnetic energy. This behavior is consistent with the results shown in Ref. \cite{JAP112:083514:2012}.

\section{Final Remarks}

We have solved the Schr\"{o}dinger equation with a confining potential, constant magnetic and electric fields in cylindrical coordinates. The radial differential equation has been mapped into a confluent hypergeometric differential equation. The energy $E_{\rho}$ was obtained in agreement with Ref. \cite{JLTP190:200:2018}. On the other hand, the differential equation related to $z$ coordinate has been mapped into a biconfluent Heun differential equation. Using (\ref{cc1}) and (\ref{cc2}) (and not only (\ref{cc1}) as done in Ref. \cite{JLTP190:200:2018}), we have shown that $E_{z}$ can not be labeled with $n_{z}$. From this fact, we can conclude that each value of $n_{z}$ can be associated to a different physical scenario. Fixing $n_{z}=1$ we have built the correct partition function and the thermodynamic properties as specific heat and entropy have been recalculated. It is shown that our results are in agreement with those found in the literature. Finally, it can be conclude that the effects of correction on the energy spectrum are strong for the specific heat (high temperatures), when compared with the results of the Ref. \cite{JLTP190:200:2018}).

\begin{acknowledgements}
The authors are indebted to the anonymous referee for an excellent and constructive review. L. B. Castro would like to thank Professor Dr. Antonio S. de Castro for useful comments and suggestions. This work was supported in part by means of funds provided by CNPq, Brazil, Grant No. 307932/2017-6 (PQ) and No. 422755/2018-4 (UNIVERSAL), S\~{a}o Paulo Research Foundation (FAPESP), Grant No. 2018/20577-4, FAPEMA, Brazil, Grant No. UNIVERSAL-01220/18 and CAPES, Brazil. Angel E. Obispo thanks to CNPq (grant 312838/2016-6) and Secti/FAPEMA (grant FAPEMA DCR-02853/16), for financial support.
\end{acknowledgements}

%
%

\bibliographystyle{spphys}       


\end{document}